\documentclass[]{aa}  
\usepackage{graphicx}
\usepackage{txfonts}
\begin{document}
   \title{X-ray imaging of the ionisation cones in NGC~5252}

   \author{M.Dadina$^{1,2}$, M.Guainazzi$^3$, M.Cappi$^1$, S.Bianchi$^4$,
     C. Vignali$^{2}$, G. Malaguti$^{1}$, A. Comastri$^5$}

   \offprints{M.~Dadina, e-mail: dadina@iasfbo.inaf.it}

   \institute{$^1$INAF/IASF Bologna, via Gobetti 101, 40129
     Bologna, Italy \\
     $^2$Universit\`a degli Studi di Bologna, Dip. di Astronomia, via Ranzani 1,
     40127, Bologna,  Italy \\
     $^3$European Space Astronomy Center of  ESA, Apartado 50727, E-28080
     Madrid,Spain \\
     $^4$Dipartimento di Fisica, Universit\`a degli Studi Roma Tre, via della
     Vasca Navale 84, I-00046, Roma, Italy\\
     $^5$INAF/Osservatorio Astronomico di Bologna, via Ranzani 1, 40127,
     Bologna, Italy\\ 
     }
   
   \date{}

% \abstract{}{}{}{}{} 
% 5 {} token are mandatory
 
  \abstract
   {The physical conditions of the gas forming the narrow line regions 
(NLR) in active galactic nuclei (AGN) have been extensively studied in the optical
band. Recently, detailed X-ray studies have shown how the emission in the 0.1-2 keV
band detected in Seyfert 2 galaxies is associated to gas lying close to or
associated with the NLR.}
   {We take advantage of the spectacular extension ($\sim$15'') of the NLR in the type II
     Seyfert galaxy NGC~5252 and of the complementary characteristics of
     $XMM$--$Newton$ and $Chandra$ to investigate the physical conditions of
     the gas in this galaxy. }
   {The X-ray data from $XMM$--$Newton$ are used to define the spectral
     properties of the ionising nuclear source. The $Chandra$ data are used to trace
     the spatial characteristics of the soft X-ray emission. This information
is then compared to the optical HST characteristics of the NLR in NGC~5252.}
   {The X-ray spectrum of the nucleus of NGC~5252 is intrinsically flat
     ($\Gamma$$\sim$1.4-1.5),  and absorbed by neutral gas with a column
     density 
       N$_{H}$$\sim$10$^{22}$  cm$^{-2}$. Below $\sim$1 keV a soft excess is 
      detected. The
     high-resolution spectrum obtained with the $XMM$--$Newton$ RGS 
shows the presence, in the 0.2-1.5 keV range, of emission lines which strongly 
indicate that the soft X-ray component is essentially due to ionised gas. 
Moreover, the soft X-ray emission is spatially resolved around the nucleus
     and well overlaps the images obtained in narrow optical bands centered
     around the [OIII] emission line at 5007$\AA$.  The [OIII]/soft-X
     flux ratios along the ionisation cones is basically constant.
     This indicates that the electron density does not significantly
     deviates from the r$^{-2}$ law (constant ionisation parameter)
moving outward from the nucleus.} 
   { This result combined with previous optical studies suggest
       two plausible but different scenarios in the reconstruction of the
     last $\sim$30000 years
   history of the central AGN. The most promising one is that the
   source is indeed a ``quasar relic'' with steady and inefficient energy release
   from the accretion of matter onto the central super-massive
   black-hole. This scenario is suggested also by the flat nuclear X-ray
   spectrum that suggests an advection dominate accretion flow (ADAF) like emission mechanism.}

   \keywords{Galaxies:active -- Galaxies:individual:NGC~5252 -- Galaxies:Seyfert -- X-ray:galaxies}
               
	\authorrunning{Dadina et al.}

	\titlerunning{X-ray imaging of the ionisation cones in NGC~5252}
   \maketitle
%
%________________________________________________________________

\section{Introduction}

Obscuration of the nuclear 
emission in type~II AGN allows the study of soft X-ray spectral 
components, which are normally outshone by the direct component 
in type~I unobscured objects. It 
has been well known since the early day of X-ray spectroscopy that 
excess emission above the extrapolation of the absorbed nuclear 
radiation is present in almost all bright Seyfert~2s (Turner et al. 
1997). This excess appears smooth when measured with instruments with moderate
energy resolutions such as CCD. 
However, high-resolution (grating) measurements with {\it Chandra} and 
XMM-Newton revealed that this excess is generally due to a blending of strong 
recombination lines from He- and H-like transitions of elements from 
Carbon to Nitrogen (Sako et al. 2000, Sambruna et al. 2001, Kinkhabwala 
et al. 2002, Armentrout et al. 2007). X-ray spectral diagnostics 
(Kinkhabwala et al. 2002, Guainazzi \& Bianchi 2006) and a close 
morphological coincidence between the soft X-rays and the [OIII] in 
Extended Narrow Line Regions (ENLR; Bianchi et al. 2006, Bianchi et al. 2010) strongly 
indicate that the gas is photoionised by the AGN, with an important role 
played by resonant scattering.

In this context, NGC~5252 represents an extraordinary laboratory to study the 
feedback between the AGN output and circumnuclear gas on kpc scale, thanks to 
its spectacular ionisation cones (Tadhunter \& Tsvetanov 1989).

NGC 5252 is classified as Seyfert 1.9 
(\cite{ost}) S0 (\cite{dev}) nearby (z=0.023,) galaxy (N$_{H,
  Gal}$=2.14$\times$10$^{+20}$ cm$^{-2}$, Dickey \& Lockman, 1990). Small
radio jets (r$\sim$4'') have
been detected and found to be aligned with the ionisation cones
(\cite{wilson94}). Nonetheless, the host galaxy luminosity
(M$_{R}$$\sim$-22. \cite{capetti}), 
mass (M$_{bulge}$$\sim$2.4$\times$10$^{11}$M$_{\sun}$, \cite{marc}) and the
mass of the central super-massive black-hole
(M$_{BH}$$\sim$10$^{9}$M$_{\sun}$, \cite{capetti}) are more typical of 
quasar than Seyfert galaxies. These pieces of evidence led
\cite{capetti} to speculate
that NGC 5252 is most probably to be considered a QSO relic. This view is in
agreement with "downsizing" scenarios about the evolution of super-massive
black-hole (SMBH) in cosmic
times. Accordingly with these scenarios, most massive SMBHs formed and evolved
earlier than lower mass ones.

Ionisation cones are one of the strongest argument in favour of the Seyfert 
unification scenarios (Antonucci 1993). For this reason, NGC~5252 is also an 
important laboratory to test AGN geometrical models. From a diferent point of
view, 
AGN activity has been recognized, since a 
while as a key component of the SMBH host galaxy co-evolution and AGN
feedback is likely to self-regulate or be responsible of the observed 
properties (Menci et al. 2004).  
The very existence of ionization cones witness that feedback/winds
   are or were active and thus these sources are ideal laboratories
   for feedback.

X-ray measurements allows to directly link the 
properties of the gas emitting optical lines with the intrinsic AGN 
power, which in type~II AGN can be truly measured only at energies 
larger than the soft photoelectric cut-off due to the AGN obscuring 
matter. Furthermore, the morphological coincidence between X-rays and 
optical emission in ENLR (Bianchi et al. 2006) points to a fundamental 
physical link between the two wavebands. They need to be studied 
simultaneously in order to derive the correct energy budget in the 
ionisation cones. Prompted by these motivations, we have performed deep 
X-ray observations of NGC~5252 at the highest spatial and spectral 
resolution currently available with {\it Chandra} and XMM-Newton. The 
results of these observations are the subject of this paper.

\section{The nuclear spectrum}

NGC5252 was observed by XMM-Newton on 2003, July 18th, 
with the EPIC CCDs (MOS and pn in full window, see Tab.~\ref{tab11}) as the prime instrument
for a total duration of $\sim$67 ks. The
{\it Observation Data Files} (ODFs) were reduced and analysed using the 
latest Science Analysis System (SAS) software package 
(\cite{gabriel03}) with associated latest 
calibration files. We used the {\tt epproc} and {\tt emproc} tasks to 
generate event files and remove dead and hot pixels. 
Several time intervals with a high background rate were 
identified in the
single events light curve at energy $>$10 keV and were 
removed, yielding a net exposure of $\sim$50 ks for the MOS
and $\sim$38 ks for the pn. Pile-up is negligible in this source,
according to the {\tt epatplot} SAS task outcome.
Patterns $\leq$12 and $\leq$4 were considered for MOS and 
pn, respectively. Source counts were extracted from a circular region
with radius 50$\arcsec$, thus encompassing a large
fraction of the optically-defined galaxy. Background was estimated using both
blank-sky files and locally from a offset source-free region.
Light curves in the soft (0.5-2 keV) and hard (2-10 keV) energy bands 
were first investigated. We found no significant flux nor spectral 
variations, thus considered the time averaged spectrum.

\begin{figure}
   \centering
   \includegraphics[angle=-90,width=8.0cm]{spe_cont.ps}
\vspace{-2.0mm} \includegraphics[angle=90,width=8.0cm]{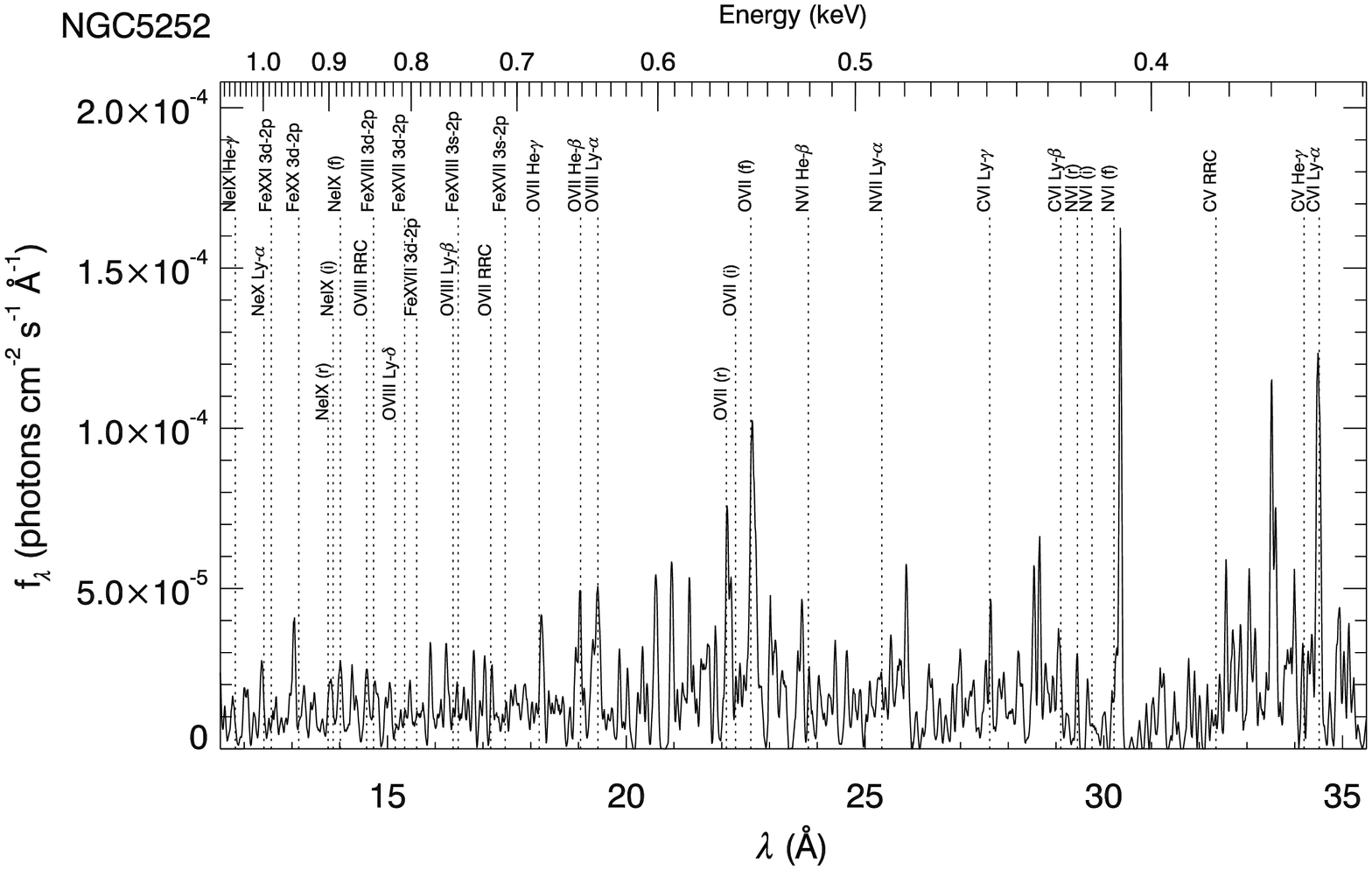}
%\vspace{-2.0mm}   \includegraphics[width=8.0cm]{fig1_rgs.ps2}
      \caption{{\it Upper panel}: XMM-Newton EPIC spectrum extracted from a
        50$\arcsec$ region around the NGC~5252 nucleus. For clarity, only pn
        data are presented.  {\it Lower panel:} XMM-Newton RGS spectrum of
        NGC~5252 }
         \label{fig1mc}
   \end{figure}

The best-fit spectrum is shown in Fig.~\ref{fig1mc}.
%------------------------ Figure 1 MC
%------------------------ Figure 1
If the nuclear emission is modeled with a simple absorbed power-law we
obtain an extremely flat photon index [$\Gamma$=1.05$\pm$0.10 and
$N_{\rm H}$=(2.2$\pm$0.1)$\times 10^{22}$~cm$^{-2}$, the reported errors
  are, here and hereafter, at 90\% confidence level], 
plus a soft component emerging at energies below E$\sim$1 keV. It can be
parametrised with a scattered power-law with a steep photon index $\Gamma$=3.0$\pm$0.2.
There is also evidence for
one (or few) soft emission lines in addition to the soft
continuum, a Fe K$_{\alpha}$ line with 
E$=$6.44$\pm$0.05 keV and EW=50$\pm$25 eV, and an absorption edge 
at E=7.0$\pm$0.1 keV and optical depth $\tau$=0.31$\pm$0.05.
 A significantly better fit  
is obtained in case of a power-law plus a thermal
  component, namely $mekal$ in $Xspec$ (\cite{mewe85}), 
    is used to model the soft X-ray band
of NGC 5252 ($\chi^{2}$/d.o.f.=1436/1410 in the first case while
$\chi^{2}$/d.o.f.=1329/1410 with $mekal$). 
In this case, the temperature of the plasma
is $kT$$\sim$0.17 keV. 
The investigation on the true nature of
this soft X-ray component will be the subject of the next Sections.  
 Here it is important to note that, whatever the fitting of the data below $\sim$1 keV, the best-fit
model for the nuclear emission of  NGC 5252 above $\sim$1 keV
is typical in shape, but flatter ($\Gamma$$\sim$1.4-1.5) than normally
found from Seyfert 2 galaxies ($\Gamma$$=$1.5-2.5, Turner \& Pounds 1989; \cite{turner97,risaliti02,cappi06}; Dadina 2008). It is however 
consistent with the previous ASCA measurement ($\Gamma$$\sim$1-1.5,
\cite{cappi96}), confirming the need in this source for a more complex 
absorption  
(either multiple, ionised or both) in order to recover a steeper 
canonical photon index.
With the above model we measure, for the soft (0.5-2 keV) component, a flux of 
3.5 $\times$10$^{-13}$erg cm$^{-2}$s$^{-1}$ corresponding to a luminosity of 
4.1$\times$10$^{41}$ erg/s and, for the hard (2-10 keV) component, a flux of 
8.9 $\times$10$^{-12}$erg cm$^{-2}$s$^{-1}$ corresponding to a (unabsorbed) 
luminosity of 1.2$\times$10$^{43}$ erg/s. This is consistent, within a few tens
percent, with previous ASCA and BeppoSAX values (Cappi et al. 1996; Dadina
2007). It is worth noting here that,
from IR diagnostics, we expect that star forming activity should contribute to
less than $\sim$1\% to the total soft X-ray emission (Cappi et al. 1996).

\section{High-resolution spectroscopy of the AGN environment}
%------------------------ Figure 2
   \begin{figure*}
   \centering
   \includegraphics[width=8cm,angle=-90]{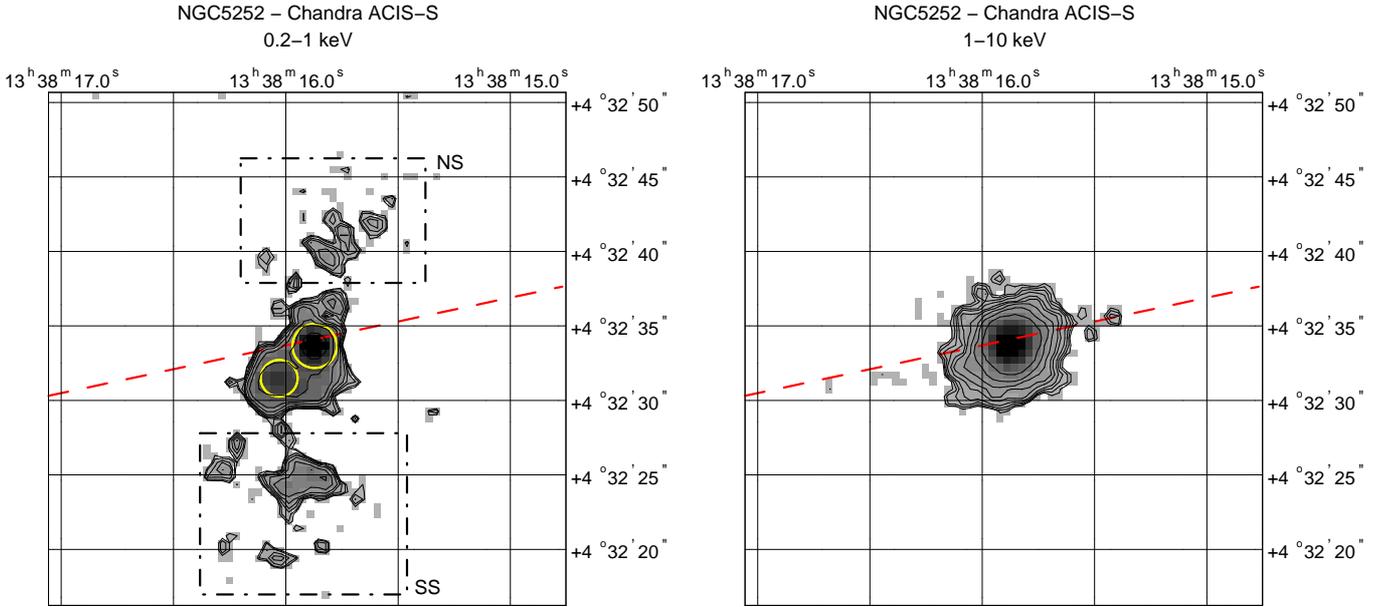}
      \caption{$Chandra$ ACIS-S images of NGC~5252 in the 0.1--1~keV (soft;
 		{\it left panel}), and 1--10~keV (hard; {\it right
		panel}) Images were smoothed with a
		$\sigma = 1.25$~pixels (yielding an angular resolution of
                  $\simeq$1" in the images)
		wavelet for illustration purposes. The {\it
		thin solid lines} represent
		9 linearly spaced contours in the range 2 to 20 counts per
		pixel. The {\it thick dot-dashed line} indicate the position
		of the out-of-time events readout streak, removed before
		the generation of the image. The {\it dashed} lines represents
		the regions, whence the spectra
		of the South and North diffuse Spots were
		extracted. The {\it solid circles} represent the regions,
		whence the spectra of the nucleus (the brightest and
                  central spot) and of the South-East
		Nuclear Source were extracted.
              }
         \label{fig2}
   \end{figure*}
%------------------------ Figure 2

The Reflection Grating Spectrometer (RGS; \cite{derherder01})
on board XMM-Newton
observed NGC~5252 simultaneously to the EPIC cameras (Tab.~\ref{tab11}, Fig.~\ref{fig1mc}). It
produces high-resolution (first order resolution 600-1700~km~s$^{-1}$)
spectra
in the 6--35~\AA\ (0.35--2~keV) range. Its 2.5$\arcmin$ diameter
slit fully encompasses the ionisation cones and the host galaxy.
The RGS spectrum represents therefore only the average conditions of
the soft X-ray emitting gas across the nucleus and the cone.

\begin{footnotesize}     
\begin{table}
\caption{Main characteristics of the X-ray observations presented here.} 
\label{tab11}
\centering          
\begin{tabular}{l c c l c c }     % 5 columns 
\hline
&&&&&\\
Instr.&Exp.&CR & Instr.&Exp.&CR \\
&&&&&\\
&ks&c/s&&ks&c/s\\
&&&&&\\
\hline\hline       
&&&&&\\
Epn&38&1.10$^{a}$& RGS1&63& 0.08$^{b}$\\
&&&&&\\
EMOS1&49&0.37$^{a}$ &RGS2&63& 0.08$^{b}$\\
&&&&&\\
EMOS2&50&0.37$^{a}$&ACIS-S&60&0.32$^{a}$\\
&&&&&\\
\hline
\end{tabular}

$^{a}$ count-rate in the 0.2-10 keV band; $^{b}$ count-rate in the 0.4-1.2 keV
\end{table}
\end{footnotesize}

RGS data were reduced starting from the {\it Observation Data Files}
with SASv6.5 (\cite{gabriel03}), and using the latest calibration files. 
The SAS meta-task {\tt rgsproc} was used
to generate source and background spectra, assuming as a reference
coordinate coincident with the optical nucleus of NGC~5252. Background
spectra were generated using both blank field maps - accumulated across
the whole mission - and a ``local'' background accumulated during the
observation. The former, based on a model of the estimated background on
the basis of the count rates detected in the most external of the camera
CCDs, overestimates the intrinsic background level during the
observation. We have therefore employed the ``local'' background hereafter.
A correction 
factor to the count background spectrum has been applied to take into 
account the size of the extraction region, which corresponds to the area 
of the RGS active CCDs outside the 98\% percentage point of the line 
spread function in the cross-dispersion direction.

%------------------------ Figure 3
   \begin{figure}
   \centering
   \includegraphics[width=8cm]{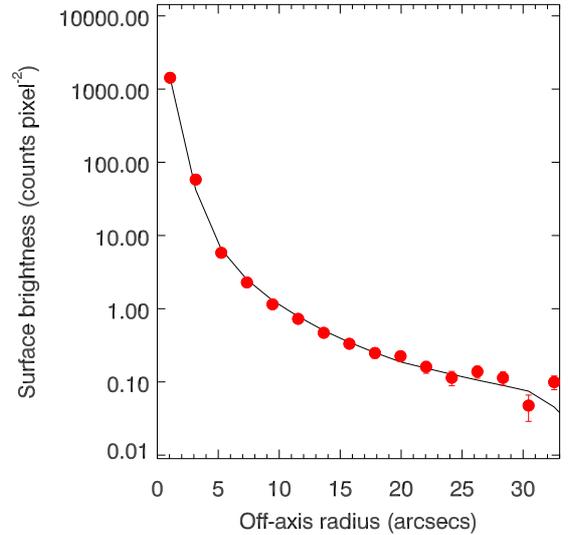}
      \caption{Radial profile ({\it filled circles}) of the ACIS-S hard band
		image. The {\it solid line}
		represents the PSF for a source with the same
		hard X-ray spectral energy distribution as the NGC~5252
		``nucleus normalized'' to its on-axis peak flux. When not visible,
                the error bars are within the filled circles.
              }
         \label{fig3}
   \end{figure}
%------------------------ Figure 3

We simultaneously fit the spectra of the two cameras 
following the procedure outlined in Guainazzi \& Bianchi
(2006)\footnote{This paper discusses a sample of 69 RGS spectra of
type 1.5, 1.8, 1.9 and 2 Seyfert galaxies. 
The observation of NGC~5252 discussed in this paper belongs to this sample as well.}
who have performed local spectral fits around
each of the $\simeq$40 emission lines detected in the archetypal
obscured Seyfert NGC~1068 (\cite{kin02}). In these fits, both the
background level and the continuum have been assumed as independent power-law
components, with photon index $\Gamma$ set equal to 1.  
It is worth noting that the adopted 
value of the power law index, here equal to the photon index of the
continuum of the primary emission, does not affect the results signifcantly, given the very
limited band of these fits. Different choices for the 
continuum spectral index yield indistinguishable results.
Each emission line has been modeled with an unresolved Gaussian profile 
fixed
to be at the expected energies (leaving the intrinsic
width of the profile free yields a negligible improvement in the quality of
the fit). We detect three lines (see lower panel of Fig. 1) at a confidence level larger than
90\%  [$\Delta \chi^{2} = $ 10.5, 24.0, 10.8 for CV, OVII and OVIII 
lines, respectively, for one interesting parameter (Tab.~\ref{tab1})].
%-------------------- Table 2
\begin{table}
\caption{List of emission lines detected in the RGS spectrum of NGC~5252.
$E_c$ is the centroid line energy; $L$ is the intrinsic line luminosity:
$\Delta v$ is the difference between the measured line centroid energy
and the laboratory energy. Only statistical error on this measurements
are quoted. $v_{{\rm sys}}$ is the systematic error on
$\Delta v$ due to residual uncertainties in the RGS aspect solution
($\simeq$8~m\AA)} 
\begin{footnotesize}     
\label{tab1}      
\centering          
\begin{tabular}{l c c c c}     % 5 columns 
\hline\hline       
Identification & $E_c$ & $L$ & $\Delta v$ & $v_{sys}$ \\
& (eV) & (10$^{40}$~erg~s$^{-1}$) & (km~s$^{-1}$) & (km~s$^{-1}$) \\
\hline                    
C{\sc vi} Ly-${\alpha}$ & $367 \pm 5$ & $3 \pm^{11}_{2}$ & $800 \pm 400$ & 70 \\
O{\sc vii} He-${\alpha}$ (f) & $560.4 \pm 0.2$ & $2.0 \pm^{0.8}_{0.9}$ & $-370 \pm 110$ & 110 \\
O{\sc vii} He-${\alpha}$ (i) & $E_c (f) + 7.7$ & $<$0.9 & & \\
O{\sc vii} He-${\alpha}$ (r) & $E_c (f) + 13.0$ & $<$1.4 & & \\
O{\sc viii} Ly-${\alpha}$ & $654.0 \pm^{0.7}_{1.2}$ & $0.6 \pm 0.4$ & $300 \pm^{300}_{600}$ & 130 \\
\hline                  
\end{tabular}
\end{footnotesize}
\end{table}
%-------------------- Table 1
None of them is a Radiative Recombination Continuum (RRC). A (admittedly
loose) constrain on the width of the Gaussian profile can be obtained on
the O{\sc vii} He-$\alpha$ triplet only: $\sigma < 4400$~km~s$^{-1}$
(8.2~eV).
%------------------------ Figure 4
   \begin{figure*}
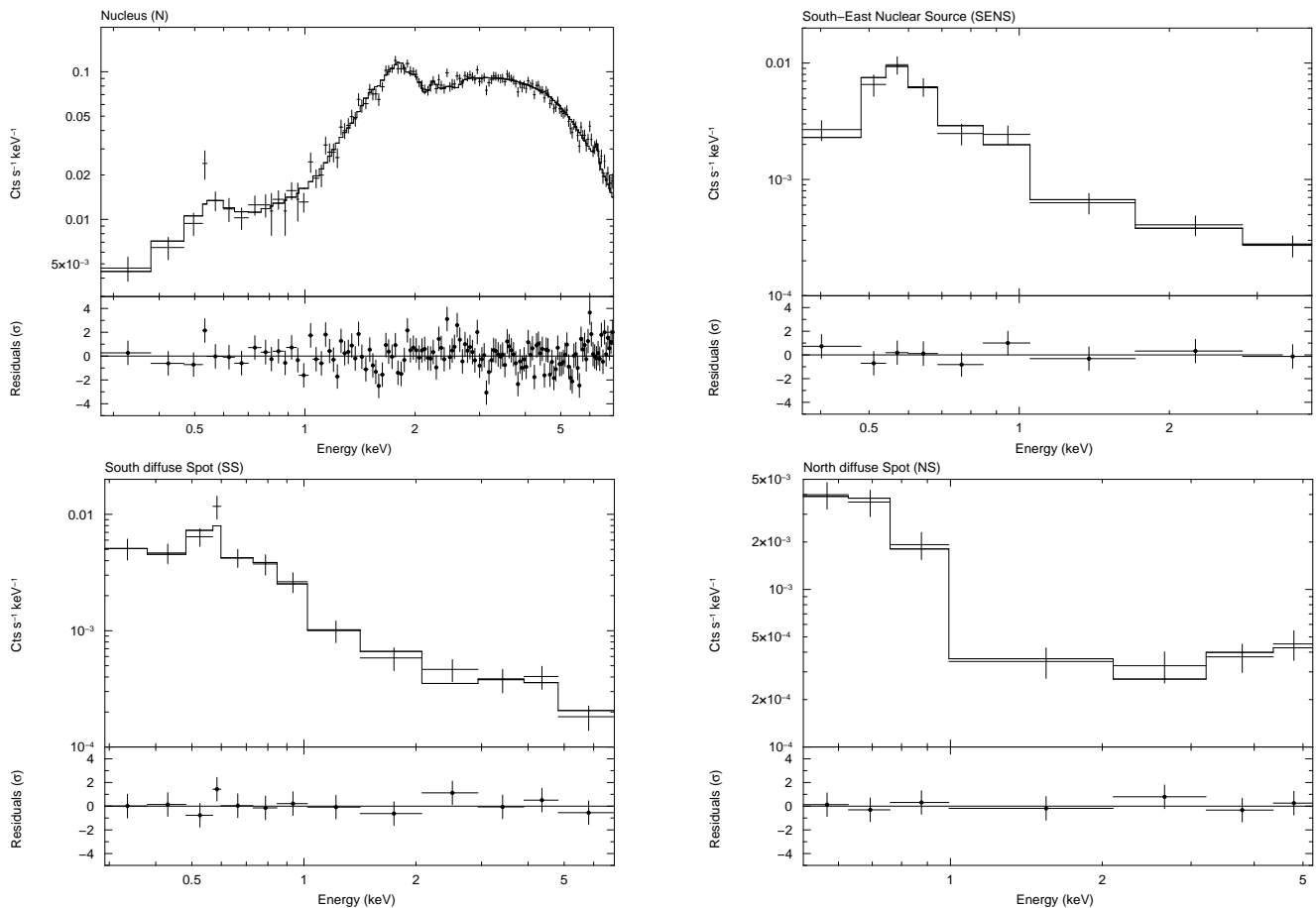

   \centering
   \hbox{
   \includegraphics[width=6cm,angle=-90]{fig4a_1.ps}
   \hspace{1.0cm}
   \includegraphics[width=6cm,angle=-90]{fig4b_1.ps}
   }
   \hbox{
   \includegraphics[width=6cm,angle=-90]{fig4c_1.ps}
   \hspace{1.0cm}
   \includegraphics[width=6cm,angle=-90]{fig4d_1.ps}
   }
      \caption{Spectra ({\it upper panels}) and residuals in units of
		standard deviations ({\it lower panels}) for the four
		regions of NGC~5252 defined in Fig.~\ref{fig2}.
              }
         \label{fig4}
   \end{figure*}
%------------------------ Figure 4
Diagnostic parameters involving the intensity of the O{\sc vii} He-${\alpha}$
triplets can be in principle used to pinpoint the physical process responsible
for the bulk of the X-ray emission in high-resolution spectra. The detection
of the forbidden ($f$) component only allows to set lower limits
on the standard triplet diagnostics (\cite{gabriel69,porquet00}):
$R > 1.1$, $G > 0.7$ (where R is the ratio between forbidden and
intercobination lines and depends on the electron density, while G is the
ratio between intercombination plus forbidden lines and the resonance line, \cite{gabriel69,porquet00}). 

These limits, although fully consistent with
photoionised plasmas, do not rule out collisional ionisation. Guainazzi \&
Bianchi (2007) proposed a criterion to discriminate, {\it on a statistical
basis}, between AGN- and starburst-powered sources based on the location
of the source in an empirical observable plane: integrated luminosity of the
He- and H-like Oxygen lines, $L_O$, against the intensity
ratio $\eta$ between the $f$ and the
O{\sc viii} Ly-$\alpha$. In NGC~5252
$\eta = 2.3 \pm 0.4$, and
$L_0 \sim 3 \times 10^{40}$~erg~s$^{-1}$. These values put NGC~5252 in
the plane locus preferentially occupied by photoionised (AGN) sources
(\cite{guainazzi09}).
We estimated also the flux density associated to the continuum, using a
line-free energy range between 586 and 606~eV: $\nu L_{\nu}|_{0.6 \ keV} =
(7.2 \pm^{1.7}_{2.9}) \times 10^{40}$~erg~s$^{-1}$.

\section{X-ray imaging of the ionisation cones}

{\it Chandra} observed NGC~5252 on August 11, 2003 with the ACIS-S
detector in standard VFAINT configuration. Data reduction was performed
with CIAO version 3.3 and associated CALDB version 3.2. ``Level~1''
events were corrected for bad pixels, gain spatial dependency,
and charge transfer inefficiency via {\tt acis\_process\_events}.

Although the correction for read-out streaks was applied
as well, some out-of-time events remain in the final cleaned event
list, and were removed by applying a 2 pixels ($\simeq$1$\arcsec$)-wide
tilted rectangular box around the streak.

ACIS-S images in the
$\sim$2$\arcmin$ around the optical core of
NGC~5252 are shown in Fig.~\ref{fig2} in the 0.2--1~keV  and
in the 1--10~keV energy bands. The soft band clearly shows extended emission
in the North-South direction on both sides of the nucleus. On the
contrary, the hard band image is point-like. We extracted a radial
profile of the latter, and compared it with the expected instrumental Point
Spread Function (PSF) of a source
with the same spectral energy distribution as the
NGC~5252 nucleus. The two profiles are perfectly consistent
up to 30$\arcsec$ off-axis (see Fig.~\ref{fig3}).

In order to characterize the spectral behavior of the diffuse emission,
we have extracted spectra from four regions, identified in the soft image
(Fig.~\ref{fig2}): the nucleus (N), a S-E source about 3.2$\arcsec$ from the
nucleus (SENS), and the South (SS) and North (NS) diffuse Spots. Background
spectra were generated from a large circle 57$\arcsec$ wide around the
galaxy core, once a $21$$\arcsec$ inner circle, as well as
5$\arcsec$ circles around each serendipitous point sources were removed. Alternative
choices of the background regions do not substantially change
the results presented in this section. Source spectra were rebinned
in order to over-sample the intrinsic instrumental energy resolution by
a factor $\ge$3, and to have at least 25 background-subtracted counts
in each spectral bin. The latter criterion
ensures the applicability of the $\chi^2$ statistics.

For all spectra, we have employed a baseline model that include 
a thermal emission component from collisionally excited plasma ({\tt mekal} in
{\sc Xspec}; \cite{mewe85}). 
This choice was done for simplicity and the only information
obtained using  {\tt mekal} is the flux of the thermal component. 
This is particularly true for the SS and NS regions where the complexity of 
the {\tt mekal} model is well above the quality of the data. 
Moreover, a photoelectrically-absorbed power-law was
always included in the data. 
The physical meaning of the latter is different
depending on the region where the spectrum was extracted. For the nuclear
region, the non-thermal component represents the contribution of the active
nucleus; for the other regions, the integrated contribution of hard galactic
sources such as, for example, X-ray binaries, cataclysmic variables or 
supernova remnants. We therefore
refrain from attributing a physical meaning to the power-law spectral
indeces in the latter case.
The spectra and corresponding best-fits are shown in Fig.~\ref{fig4}.
A summary of the spectral results is presented
in Tab.~\ref{tab2}. The {\it Chandra} data confirm that the nuclear spectrum
%------------------------ Table 2
\begin{table*}
\label{tab2}
\caption{ACIS-S best-fit parameters and results for the spatially-resolved regions
of NGC~5252. $E_c$ and $EW$ are the centroid energy and the Equivalent Width
of a Gaussian profile at the energies of K$_{\alpha}$ fluorescence for
neutral or mildly ionised iron. Fluxes, ($F$), are in the observed frame.
Luminosities, ($L$), are in the source frame, and are corrected for absorption.}
\begin{tabular}{lcccccccccc} \hline \hline
Region & $N_H$ & $\Gamma$ & $kT^a$ & $E_c$ & $EW$ & $F_{0.5-2 keV}$$^b$ &
$F_{2-10 keV}$$^b$ & $L_{0.5-2 keV}$$^c$ & $L_{2-10 keV}$$^c$ & $\chi^2/\nu$ \\
&  $(10^{22}$~cm$^{-2}$) & & (eV) & (keV) & (eV) & & & & & \\ \hline
Nucleus & $2.32 \pm^{0.13}_{0.15}$ & $1.00 \pm^{0.08}_{0.06}$ & $140\pm^{60}_{40}$ & $6.37 \pm^{0.06}_{0.05}$ & $50 \pm 20$ & $2.66\pm^{0.11}_{0.28}$ & $103\pm^2_3$ & $0.82\pm^{0.03}_{0.09}$ & $1130 \pm^{20}_{30}$ & 176.9/127 \\
SENS & $0.47\pm^{0.16}_{0.31}$ & $0.8 \pm 0.4$ & $<140$ & ... & ... & $0.06 \pm 0.05$ & $0.34 \pm^{0.08}_{0.14}$ & $50 \pm 40$ & $4.1 \pm^{1.0}_{1.7}$ & 3.0/5 \\
NS & $\equiv$$N_{H,Gal}$ & $-0.9\pm^{0.5}_{0.6}$ & $240 \pm^{40}_{30}$ & ... & ... & $0.066\pm^{0.011}_{0.015}$ & $1.3 \pm 0.3$ & $0.89 \pm^{0.15}_{0.20}$ & $15 \pm 3$ & 1.0/3 \\
SS & $\equiv$$N_{H,Gal}$ & $0.0 \pm 0.4$ & $110\pm^{13}_{21}$ & ... & ... & $0.122 \pm^{0.017}_{0.023}$ & $0.70 \pm^{0.18}_{0.53}$ & $1.7 \pm^{0.2}_{0.3}$ & $8 \pm^2_6$ & 5.1/7 \\
& & & $510\pm^{170}_{230}$ & & & &  &  & &  \\ \hline
\end{tabular}

\noindent
$^a$derived using $mekal$ to fit the spatially resolved data

\noindent
$^b$in units of $10^{-13}$~erg~s$^{-1}$~cm$^{-2}$

\noindent
$^c$in units of $10^{40}$~erg~s$^{-1}$

\end{table*}
%------------------------ Table 2
is remarkably flat, and is
seen through a substantial column density ($N_H \simeq
2.26 \times 10^{22}$~cm$^{-2}$).

\begin{figure}
   \centering
   \includegraphics[width=8.5cm]{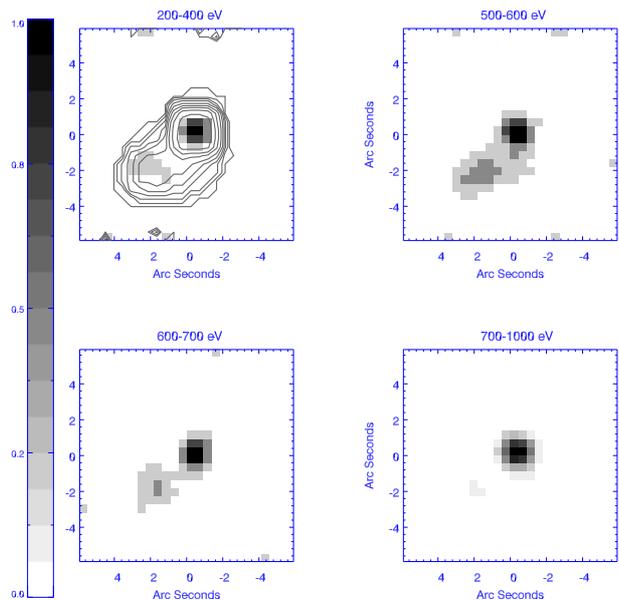}
      \caption{Narrow-band ACIS-S images in the 10$\arcsec$ around the
		NGC~5252 core, normalized to the peak nuclear emission.
		Images are smoothed with a 5 pixel Gaussian kernel.
		The {\it solid lines} in the {\it upper left} panel
		represent a contour plot of the 0.2--1~keV 
		image, assuming the same smoothing criterion.
              }
         \label{fig5}
   \end{figure}

The background subtraction for the spectrum of SENS could be contaminated
by the spilling of the nuclear emission.
The encircled energy fraction at a distance equal to that between
sources N and SENS is $\simeq$97.5\%. However,
subtracting a properly rescaled nuclear spectrum to the SENS spectrum yields
negative counts above 2~keV. In order to have an independent estimate of the
spectral behavior of SENS, we have extracted images
10$\arcsec$ around the nuclear region in narrow energy bands
(Fig.~\ref{fig5}): 200-400~eV,
%------------------------ Figure 5
%------------------------ Figure 5
500-600~eV, 600-700~eV and 700-1000~eV. In a line-dominated plasma, the
above energy ranges correspond to bands dominated by C{\sc vi}
and C{\sc v} K$_{\alpha}$, O{\sc vii} He-$\alpha$, O{\sc viii} Ly-$\alpha$,
and Fe-L transitions, respectively.
Each image was normalized to the peak of the
nuclear emission in that energy band. The soft X-ray
SENS spectrum is comparatively dominated by Oxygen transitions, with little
contribution in either the Carbon or the Iron band.

\section{Comparing soft X-ray and [OIII] morphologies}

NGC 5252 was observed in the [OIII] band with the WFPC2 on-board HST on
1995, July 23, using the linear ramp filter FR533N. The data were downloaded
from MAST (multi-mission archive at STScI). The images were processed through the standard OTFR (on-the-fly reprocessing) calibration pipeline which performs analog-to-digital conversion, bad pixel masking, bias and dark subtraction, flat field correction and photometric calibration. The cosmic rays rejection was performed combining the two images that are usually taken for this scope. Geometric distortion was corrected using the {\it multi drizzle} script (\cite{koekemoer02}).

The relative Chandra-HST astrometry is clearly a fundamental issue for this
work. Chandra has a nominal position accuracy of 0.6$\arcsec$ while the
absolute astrometry of HST is accurate to 1-2$\arcsec$. Fortunately, to align
the two astrometric solutions, we could use a point-like source detected both
in the WFPC2 and Chandra fields. This source was previously detected at radio
wavelengths (\cite{wilson94}) and is most probably associated to a background
quasar (Tsvetanov et al. 1996). Moreover, as a second reference point, we used the brightest emission peak in the nuclear region of NGC 5252 itself.

Images were calibrated in flux using a constant flux conversion
factor of $1.839 \times 10^{-16}$, corresponding to the flux producing a
count rate of 1~s$^{-1}$ in the filter band. The above value is appropriate
for the instrumental configuration employed during the 
NGC~5252 exposure,
as indicated by the {\tt PHOTFLAM} keyword in the image file.

\begin{figure}
    \centering
     \includegraphics[width=8.5cm]{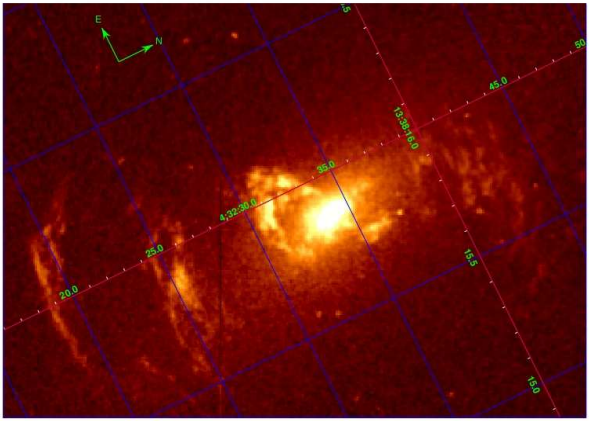}
       \caption{HST WFC~2 [OIII] image of the ionisation cones in NGC~5252.
               }
          \label{fig6}
    \end{figure}

The HST image is shown in Fig.~\ref{fig6}.
%------------------------ Figure 6
%------------------------ Figure 7
Tsvetanov et al. (1996), Morse et al. (1998) \& Capetti et al. (2005)
discussed it in details.
We refer the reader to these papers for an extensive
discussion on the properties of the optical emission. Their main
outcomes can be summarized as follows:

\begin{itemize}

\item the surface brightness is dominated by the unresolved nucleus

\item a half-ring structure is apparent S-E of the nucleus at a
maximum projected distance of $\simeq$1.5~kpc. It is probably associated
with the near side of an inclined gas disk, whose far side is
obscured by the host galaxy dust (\cite{morse98});

\item the large scale ionisation cone is traced by thin shells of
enhanced emission at either side of the nucleus, well aligned along
a P.A.$\simeq$110$^{\circ}$ at distances between 5 and 11~kpc. Fainter
co-aligned structures at scales as large as 20~kpc are 
detected as well in the O[{\sc iii}] images;
however, we will not discuss these latter structures, as
they are beyond the region where X-ray emission associated with
NGC~5252 is detected;

\item there is no evidence of radial motions. The measured velocities
of the different structures are fully explained by the rotations of the 
two disks [nonetheless Acosta-Pulido et al. (1996) claimed the detection
of radial motions describing the kinematic properties of the [OIII] emission 
arcs].

\end{itemize}

\begin{figure}
    \centering
    \includegraphics[width=8.5cm,angle=-90]{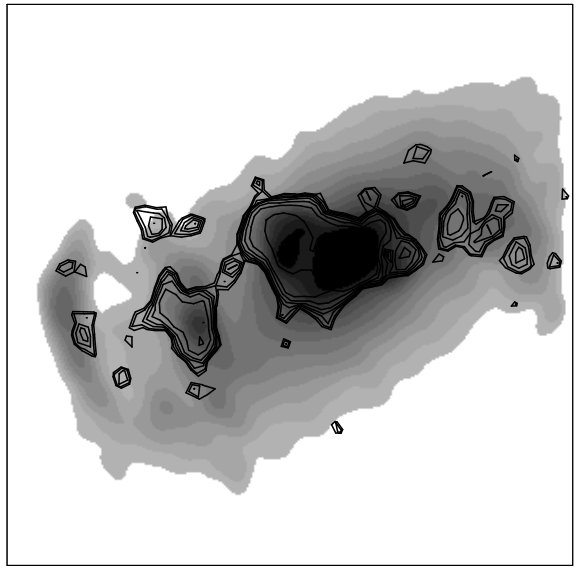}
       \caption{Iso-intensity {\it Chandra}-ACIS 0.2-1 keV X-ray iso-intensity
 		contours superposed to the HST WFC~2 [OIII] image
 		of Fig.~\ref{fig7}. The resolution of the latter has been
 		degraded to the typical resolution of {\it Chandra} optics
 		by applying a wavelet smoothing with an 8~pixel
 		kernel. The {\it Chandra} contours represents
 		nine linearly spaces count levels from 0.5 to 20 counts
 		per pixel, after a wavelet smoothing with a $\sigma$=1.25
 		pixel has been applied.
               }
          \label{fig7}
    \end{figure}

In Fig.~\ref{fig7} we present
 %------------------------ Figure 7
% ------------------------ Figure 7
the superposition between the soft X-rays iso-intensity contours to
the [OIII] image. The HST image spatial resolution has been
degraded with a 8~pixel wavelet kernel to match the resolution
of the {\it Chandra} optics.
Regions of enhanced X-ray emission exhibit a remarkable coincidence with the
morphology of the optical narrow-band image. 

 %------------------------ Figure 8
    \begin{figure}
    \centering
    \includegraphics[width=5.5cm,angle=-90]{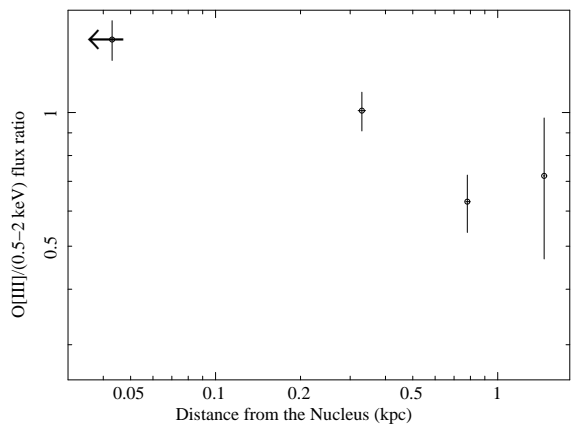}
       \caption{[OIII]/Soft-X flux ratio as a function of distance from the nucleus}
          \label{fig8}
    \end{figure}

We have calculated the ratio between the
[OIII] band and the 0.5-2 keV flux (Fig. 8) for the regions specified, 
after splitting region SS into two
sub-regions divided by a E-W line at $\delta_{J2000} = 4^{\circ} 32\arcmin
21\arcsec$ (regions ``SSNorth'' - SSN - and ``SSSouth'' - SSS - respectively).
The ratio exhibit a dynamical range smaller than a factor 2
over distances ranging from less than 100 pc
to $\sim$1.5 kpc with a slight tendency to decrease with the distance from the 
nucleus(r). This last effect, however, is most probably an observational 
artifact due to the decreasing in surface brightness of the arcs moving away
from the nucleus coupled with the sensitivity limits of $Chandra$ to extended
sources. At a first glance, the [OIII]/Soft-X ratio profile as a function of the 
distance from the nucleus seems to 
suggest that the electron density follow a r$^{-2}$ relation since the number 
of ionising photons and of the overall average ionisation state of the 
nuclear species remain almost constant.
This result is in 
agreement with Bianchi et al. (2006), who assumed, however,
a very simplified geometry of the emitting gas.   
A more detailed investigation on this
topic in NGC~5252 is hampered by the quality of the data.

\section{Discussion}

 The soft X-ray emission of NGC~5252 is clearly
  extended and ACIS images demonstrate that 
the spectacular ionisazion cones observed in [OIII] have counterparts
in the 0.1-1 keV band. 
The cumulative soft X-ray spectrum observed by $XMM$--$Newton$ is described by
a soft power-law ($\Gamma$$\sim$3). The ACIS images suggest that this is probably due 
to a blend of emission lines that mimics such steep power-law as demonstrated 
in other type II Seyferts like NGC 1068, Circinus galaxy and Mrk 3 (\cite{kin02}, \cite{brink02},
\cite{ogle03},  \cite{sam01}, \cite{sako00}, \cite{b05}, \cite{pp05}).
This scenario is supported also by the detection in the RGS high resolution 
spectrum of three emission 
lines, of CV, OVII and OVIII, probably due to photoionised gas. This is consistent also by previous 
optical studies that excluded 
collisional ionisation along the cones of NGC~5252 (Tsvetanov et al. 1996). 
Moreover, the presence of {\it in situ} ionisation sources due to shocks formed by 
large scale outflows interacting with the interstellar matter has been
excluded (\cite{morse98}). 
Thus the source of ionising photons is most probably the nucleus. Under this
assumption,  
we can use the imaging of the arcs to study the physical condition
of the gas along the ionisation cones. In particular, 
the constant of [OIII]/(0.5-2 keV) flux ratio along the ionisation cones 
within the inner\footnote{The outer arcs and filaments (\cite{tad}) are most probably too weak 
to be detected in X-rays. Considering the extension of the outer [OIII] arcs, the
minimum detectable flux between 0.1-1 keV is F$_{0.1-1
  keV}$$\sim$5$\times$10$^{-15}$erg s$^{-1}$ cm$^{-2}$ 
while, assuming
a constant [OIII]/soft X-ray ratio, the expected flux should be $\sim$10 times
lower. } $\sim$1.5 kpc suggests a r$^{-2}$ law
for the ion density.

%It is worth considering here, in fact, that $Chandra$ allows to detect of 
%the arcs that are within 15'' from the 
%nucleus.  The outer arcs and filaments (\cite{tad}) are most probably too weak 
%to be detected in X-rays. Considering the extension of the outer arcs, the
%expected minimum detectable flux should be  
%F$_{0.1-1 keV}$$\sim$5$\times$10$^{-15}$erg s$^{-1}$ cm$^{-2}$ while, assuming
%a constant [OIII]/soft X-ray ratio, the expected arcs flux should be $\sim$10 times
%lower between 0.1 and 1 keV. 

%U\footnote{U=$\frac{L}{4 \pi \rho r^{2}}$, where L is the source's luminosity,
%  $\rho$ is the density of the ionized gas, $r$ is the distance between the
%  source of ionizing photons and the ionized matter.} 
%is constant along the
%cones. This results is partially in agreement with optical studies
%(\cite{ap96}) that showed that U$\propto$r$^{-0.4}$ in
%the south-east ionisation cone, while U is constant in the north-west one. 

%This information may be used with the  
%ionisation parameter U\footnote{U=$\frac{L}{4 \pi \rho r^{2}}$, where L is the source's luminosity,
%  $\rho$ is the density of the ionized gas, $r$ is the distance between the
%  source of ionizing photons and the ionized matter.} to reconstruct the history of the AGN activity 
%in the last 10$^{3-4}$ years. Optical studies (\cite{ap96}) suggested two
%different laws for U along the cones of NGC~5252, namely, U$\propto$r$^{-0.4}$ in the  
%south-east cone U=constant along the north-west cone.  

%Thus we are left with two quite different scenarios: 

Optical spectroscopic studies (Acosta-Pulido et al. 1996) suggest that the
radial dependency of the ionization parameter\footnote{U=$\frac{L}{4 \pi \rho
    r^{2}}$, where L is the source's luminosity,
  $\rho$ is the density of the ionized gas, $r$ is the distance between the
  source of ionizing photons and the ionized matter.} U follows a different law
in the south-east ($U \propto r^{-0.4}$) with respect to the north-east
($U \propto r^0$) cone. The authors speculate that the intrinsic behavior should be the one
shown in the former, while the radial-independence of the ionization
parameter in the latter may be due to a ``conspiracy'' introduced by the
existence of two counterotating disks 
of gas (\cite{morse98}):  one is coplanar 
to the stellar one, and another is inclined by $\sim$40$^{\circ}$. 
%The morphology of the ENLR in 
%NGC 5252, thus, is supposed to be due to the superimposition of different 
%regions/rings belonging to the two disks. The  probable differences in the 
%gas distribution could have introduced the different 
%behavior of U between the southeast and northwest cones. 
Morse et al. (1998) speculated that the prominence of the southeast [OIII] 
cone in the nuclear regions is due to the fact that this component is 
seen directly, while the northeast [OIII] cone is seen
through the gas of the other disk.  
If so, the absorption due to this component could alter the line ratios 
presented by \cite{ap96} and thus the correct behavior of U should be the one derived from
the south-east cone. U$\propto$r$^{-0.4}$ implies that the luminosity of the nucleus increased by a
factor $\Delta$L$\sim$3-6 in the last $\Delta$t$\sim$5000 years.  These numbers
become $\Delta$L$\sim$10-30  and $\Delta$t$\sim$30000 years if we further 
assume that
the U and the ion density laws are still valid up to 10 kpc from the nucleus,
i.e. where the optical cones are still detectable in [OIII] but not in
X-rays.

On the contrary, having U constant and $\rho \propto$ r$^{-2}$ would imply that L has remained constant during the last 5000
(30000) years. This is consistent with the ``quasar-relic'' scenario proposed by
\cite{capetti}. These authors suggested that the nucleus of NGC~5252 is indeed
the ``relic'' of a nucleus that already experienced the activity phase in the
past and that now persists in an almost quiescent phase. 
This is suggested by the high mass of the SMBH  (M$_{BH}$=10$^{9}$M$_{\sun}$, 
Capetti et al. 2005) that  
indicates that the nucleus has already accreted in the past, the low 
Eddington ratio (L$\sim$10$^{-3}$L$_{Edd}$, assuming the bolometric
correction from Marconi et al. (2004),
L$_{hard-x}$$\sim$(1/22)$\times$L$_{bol}$), and the early type (S0) morphology
of the 
AGN host  galaxy. In literature it is also reported that the optical emission
line ratios in the inner 30" are typical of LINERS 
 (\cite{gon}), thus suggesting that a low efficiency
engine is acting at the nucleus of the source. 
It is worth noting that also the detection of two  counterotating
disks suggests that NGC~5252 is a "quasar-relic". These
disks are tracers of a major merging event that occurred, most probably, 
more than  10$^8$ years ago, since the stellar disk of NGC~5252 is
undisturbed. If the merging event triggered a phase of AGN 
activity  (see Jogge 2006, and references therein for a discussion on this
topic), we can expect that it lasted few/some 
$\sim$10$^{7}$ years (\cite{mar}; \cite{ste}; \cite{jac}; \cite{gon2}) 
%Thus,
%also in this framework, NGC~5252 should now be, as observed, in a low 
%accretion regime.
 after which the source has persisted in a quiescent state.

Finally, it is worth noting that the spectrum of the nucleus hosted by
NGC~5252 is confirmed to be quite flat (Cappi et al. 1996). 
As shown, if modeled with a simple  absorbed power-law its photon
index points to a very hard spectrum ($\Gamma$$\sim$1). 
The low EW ($\sim$50 eV) of the neutral (E$_{FeK\alpha}$$\sim$6.4 keV) iron line
is consistent with what expected if the FeK$\alpha$ line is
produced via transmission in the observed column
(N$_{H}$$\sim$2$\times$10$^{22}$ cm$^{-2}$, Makishima 1986) thus excluding a reflection
dominated spectrum.
To reconcile,  at least marginally, 
the hardness of the NGC~5252 nuclear spectrum, with previous results for
Seyfert galaxies ($\Gamma$$\sim$1.5-2.5; Turner \& Pounds, 
1989; Nandra \& Pounds, 1994; Smith \& Done, 1996; Dadina 2008), we must invoke
complex absorption models involving partial covering of the source and/or 
the presence of ionised absorbers along the line of sight. In this
case the spectral index becomes $\Gamma$$\sim$1.4-1.5. 
    It is interesting to note that the flat photon index 
may be a further clue suggesting that the X-rays may be  
produced in an ADAF,
Narayan  \& Yi, 1994) as expected in a ``quasar-relic''.

\begin{acknowledgements}

This paper is based on observations obtained with XMM-Newton, an ESA
science mission with instruments and contributions directly funded by
ESA Member States and the USA (NASA). MD greatfully acknowledge Barbara De
Marco for the helpful discussions. MD, MC and GM greatfully acknowledge 
ASI financial support under contract I/23/05/0. CV greatfully acknowledge 
ASI financial support under contract I/088/06/0.

\end{acknowledgements}


\begin{thebibliography}{}

\bibitem[Acosta-Pulido et al. 1996]{ap96} Acosta-Pulido, 
J.~A., Vila-Vilaro, B., Perez-Fournon, I., Wilson, A.~S., 
\& Tsvetanov, Z.~I.\ 1996, \apj, 464, 177 

\bibitem[Antonucci 1993]{ant} Antonucci, R.\ 1993, \araa, 31, 473 

%\bibitem[Baganoff et al.(2001)]{2001Natur.413...45B} Baganoff, F.~K., et 
%al.\ 2001, \nat, 413, 45 


\bibitem[Baskin  \& Laor 2005]{bl05} Baskin, A., \& Laor, A.\ 2005, \mnras, 358, 1043 

%\bibitem[Bennert et al. 2006]{ben06} Bennert, N., Jungwiert, 
%B., Komossa, S., Haas, M., \& Chini, R.\ 2006, \aap, 456, 953


\bibitem[Bianchi et al.(2010)]{ } Bianchi, S., Chiaberge, 
M., Evans, D.~A., Guainazzi, M., Baldi, R.~D., Matt, G., \& Piconcelli, E.
\ 2010, arXiv:1002.0800 


\bibitem[Bianchi et al. 2006]{bgc06} Bianchi, S., Guainazzi, 
M., \& Chiaberge, M.\ 2006, \aap, 448, 499 

\bibitem[Bianchi et~al. 2005]{b05}
 Bianchi, S., Miniutti, G., Fabian, A.~C., \& Iwasawa, K.
  2005, \mnras, 360, 380

\bibitem[Brinkman et~al. 2002]{brink02}
 Brinkman, A.~C., Kaastra, J.~S., van der Meer, R.~L.~J., et~al. 2002,
  \aap, 396, 761

\bibitem[Capetti et al.  2005]{capetti}Capetti, A., Marconi, 
A., Macchetto, D., \& Axon, D.\ 2005, \aap, 431, 465 

\bibitem[Cappi et al. 1996]{cappi96} Cappi M., Mihara T., Matsuoka M.., Hayashida K., Weaver K.A., Otani C., 1996, ApJ, 458, 149 

\bibitem[Cappi et al. 2006]{cappi06} Cappi M., Panessa F., Bassani L., et al., 2006, A\&A, 446, 459

\bibitem[Dadina(2007)]{dadina07} Dadina, M.\ 2007, \aap, 461, 1209 

\bibitem[Dadina 2008]{dadina08} Dadina M., 2008, A\&A, 485, 417

\bibitem[der Herder et al. 2001]{derherder01} der Herder J., Brinkman
A.C., Kahn S.M., et al., 2001, A\&A, 365, L7

\bibitem[de Vaucoulers et al. 1991]{dev} de Vaucouleurs, G., de Vaucouleurs,
  A., Corwin, H. G., Jr.,  Buta, R. J., Pasturel, G., \& Fouque´, P. 1991,
  Third Reference  Catalogue of Bright Galaxies (New York: Springer) Devereux, N. A., Ford,

\bibitem[Dickey \& Lockman(1990)]{1990ARA&A..28..215D} Dickey, J.~M., \& Lockman, F.~J.\ 1990, \araa, 28, 215 


%\bibitem[Ferland et al. 1998]{ferland98} Ferland G. J., Korista K.T., Verner D.A., Ferguson J.W., Kingdon J.B., Verner E.M. 1998, PASP, 110, 761

\bibitem[Gabriel \& Jordan 1969]{gabriel69} Gabriel A.H., Jordan C., 1969, MNRAS, 145, 241

\bibitem[Gabriel et al. 2003]{gabriel03} Gabriel C., Denby M., Fyfe D. J.,
  Hoar J., Ibarra A., 2003, in ASP Conf. Ser., Vol. 314 Astronomical Data
  Analysis Software and Systems XIII, eds. F. Ochsenbein, M. Allen, \&
  D. Egret (San Francisco: ASP), 759 

%\bibitem[Gilli et al. 2000]{gilli} Gilli, R., Maiolino, R., Marconi, A., Risaliti, G., Dadina, M., Weaver, K.~A., \& Colbert, E.~J.~M.\ 2000, \aap, 355, 485 

\bibitem[Goncalves et al. 1998]{gon} Goncalves, A.~C., Veron, P., \& Veron-Cetty, M.-P.\ 1998, \aap, 333, 877 

\bibitem[Gon{\c c}alves et al. 2008]{gon2} Gon{\c c}alves, T.~S., Steidel, C.~C., \& Pettini, M.\ 2008, \apj, 676, 816 

\bibitem[Guainazzi \& Bianchi 2007]{guainazzi07} Guainazzi M., Bianchi S.,
  2006, MNRAS, 374, 1290

\bibitem[Guainazzi et al. 2009]{guainazzi09} Guainazzi M., Risaliti G., Nucita A., et al., 2009, A\&A, 505, 589

\bibitem[Jakobsen et al. 2003]{jac} Jakobsen, P., Jansen, R.~A., Wagner, S., \& Reimers, D.\ 2003, \aap, 397, 891 

\bibitem[Jogee 2006]{jog} Jogee, S.\ 2006, Physics of Active Galactic Nuclei at all Scales, 693, 143 

%\bibitem[Kato et al.(2009)]{2009MNRAS.tmp.1536K} Kato, Y., Umemura, M., 
%\& Ohsuga, K.\ 2009, \mnras, 1536 


\bibitem[Kinkhabwala et al. 2002]{kin02} Kinkhabwala A., Sako M., Behar E., et al., 2002, ApJ, 575, 732 

\bibitem[Koekemoer et al. 2002]{koekemoer02} Koekemoer, A. M., Fruchter, A. S.,
  Hook, R. N., \& Hack, W. 2002, in The 2002 HST Calibration Workshop,
  ed. S. Arribas, A. Koekemoer, \& B. Whitmore, Baltimore, MD: Space Telescope
  Science Institute, 339

\bibitem[Makishima(1986)]{1986LNP...266..249M} Makishima, K.\ 1986, The 
Physics of Accretion onto Compact Objects, 266, 249 


\bibitem[Marconi \& Hunt 2003]{marc} Marconi, A., \& Hunt, L.~K.\ 2003, \apjl, 589, L21 

\bibitem[Marconi et al. 2004]{marconi04} Marconi A., Risaliti G., Gilli R., Hunt L.K., Maiolino R., Salvati M., 2004, MNRAS,  351, 169

\bibitem[Martini \& Weinberg 2001]{mar} Martini, P., \& Weinberg, D.~H.\ 2001, \apj, 547, 12 

\bibitem[McLure \& Dunlop 2001]{mc} McLure, R.~J., \& Dunlop, J.~S.\ 2001, \mnras, 327, 199 

\bibitem[Menci et al.(2004)]{2004ApJ...606...58M} Menci, N., Fiore, F., 
Perola, G.~C., \& Cavaliere, A.\ 2004, \apj, 606, 58 


\bibitem[Mewe et al. 1985]{mewe85} Mewe R., Gronenschild E.H.B.M., van der Oord G.H.J., 1985, A\&AS, 62, 197

\bibitem[Morse et al. 1998]{morse98} Morse J.A., Cecil G., Wilson A.S., Tsvetanov Z.I. 1998, ApJ, 505, 159

\bibitem[Nandra \& Pounds(1994)]{1994MNRAS.268..405N} Nandra, K., \& Pounds, K.~A.\ 1994, \mnras, 268, 405 


\bibitem[Narayan \& Yi(1994)]{1994ApJ...428L..13N} Narayan, R., \& Yi, I.\ 1994, \apjl, 428, L13 


\bibitem[Ogle et~al. 2003]{ogle03}
Ogle, P.~M., Brookings, T., Canizares, C.~R., Lee, J.~C., \&
  Marshall, H.~L. 2003, \aap, 402, 849

\bibitem[Osterbrock \& Martel 1993]{ost} Osterbrock, D.~E., \& Martel, A.\ 1993, \apj, 414, 552 

%\bibitem[Panessa et al. 2006]{panessa} Panessa, F., Bassani, 
%L., Cappi, M., Dadina, M., Barcons, X., Carrera, F.~J., Ho, L.~C., \& 
%Iwasawa, K.\ 2006, \aap, 455, 173 

\bibitem[Porquet \& Dubau 2000]{porquet00} Porquet D., Dubau J., 2000, A\&AS, 143, 495

\bibitem[Pounds \& Page 2005]{pp05}
Pounds, K.~A. \& Page, K.~L. 2005, \mnras, 360, 1123

\bibitem[Risaliti 2002]{risaliti02} Risaliti G., 2002, A\&A, 386, 379

\bibitem[Sako  et~al. 2000]{sako00}
Sako, M., Kahn, S.~M., Paerels, F., \& Liedahl, D.~A. 2000, \apjl, 543,
  L115

\bibitem[Sambruna et~al. 2001]{sam01}
Sambruna, R.~M., Brandt, W.~N., Chartas, G., {et~al.} 2001,
  \apjl, 546, L9

\bibitem[Smith \& Done(1996)]{1996MNRAS.280..355S} Smith, D.~A., \& Done, C.\ 1996, \mnras, 280, 355 


\bibitem[Steidel et al. 2002]{ste} Steidel, C.~C., Hunt, 
M.~P., Shapley, A.~E., Adelberger, K.~L., Pettini, M., Dickinson, M., 
\& Giavalisco, M.\ 2002, \apj, 576, 653 

\bibitem[Tadhunter \& Tsvetanov 1989]{tad} Tadhunter, C., 
\& Tsvetanov, Z.\ 1989, \nat, 341, 422 
\bibitem[Tsvetanov et al. (1996)]{tsvetanov96} Tsvetanov, Z.~I.,
Morse, J.~A., Wilson, A.~S., \& Cecil, G.\ 1996, ApJ, 458, 172

\bibitem[Turner et al. 1997]{turner97} Turner T.J., George I.M., Nandra K., Mushotzky R.F., 1997, ApJS 113, 23
 
\bibitem[Turner \& Pounds(1989)]{1989MNRAS.240..833T} Turner, T.~J., \& Pounds, K.~A.\ 1989, \mnras, 240, 833

\bibitem[Wilson \& Tsvetanov  1994]{wilson94} Wilson, A.~S., \&
Tsvetanov, Z.~I.\ 1994, AJ, 107, 1227

%\bibitem[Xu et al.(2006)]{2006ApJ...640..319X} Xu, Y.-D., Narayan, R., 
%Quataert, E., Yuan, F., \& Baganoff, F.~K.\ 2006, \apj, 640, 319 



\end{thebibliography}
\end{document}